\documentclass[]{raa}            
\usepackage{graphicx,times}
\usepackage{natbib}

\providecommand{\bm}[1]{\mbox{\boldmath$#1$\unboldmath}}

\begin{document}

   \title{The cooling time of white dwarfs produced from type Ia supernovae
}

 \volnopage{ {\bf 2009} Vol.\ {\bf 9} No. {\bf XX}, 000--000}
   \setcounter{page}{1}

   \author{Xiang-Cun Meng
      \inst{1}
   \and Wu-Ming Yang
      \inst{1}
   \and Zhong-Mu Li
      \inst{2,3}
   }

   \institute{School of Physics and Chemistry, Henan Polytechnic
University, Jiaozuo, 454000, China; {\it xiangcunmeng@hotmail.com}\\
        \and
             College of Physics and Electronic Information, Dali University, Dali, 671003, China
        \and
        National Astronomical Observatories, Chinese
Academy of Sciences, Beijing, 100012, China
}

\abstract{Type Ia supernovae (SNe Ia) play a key role in measuring
cosmological parameters, in which the Phillips relation is
adopted. However, the origin of the relation is still unclear.
Several parameters are suggested, e.g. the relative content of
carbon to oxygen (C/O) and the central density of the white dwarf
(WD) at ignition. These parameters are mainly determined by the
WD's initial mass and its cooling time, respectively. Using the
progenitor model developed by Meng \& Yang, we present the
distributions of the initial WD mass and the cooling time. We do
not find any correlation between these parameters. However, we
notice that the range of the WD's mass decreases, while its
average value increases with the cooling time. These results could
provide a constraint when simulating the SN Ia explosion, i.e. the
WDs with a high C/O ratio usually have a lower central density at
ignition, while those having the highest central density at
ignition generally have a lower C/O ratio. The cooling time is
mainly determined by the evolutionary age of secondaries, and the
scatter of the cooling time decreases with the evolutionary age.
Our results may indicate that WDs with a long cooling time have
more uniform properties than those with a short cooling time,
which may be helpful to explain why SNe Ia in elliptical galaxies
have a more uniform maximum luminosity than those in spiral
galaxies.
  }

   \authorrunning{Meng,  Yang \& Li}            
   \titlerunning{The cooling time of white dwarfs produced from Type Ia Supernova}  
   \maketitle


%
%
\section{INTRODUCTION}           
\label{sect:1} As one of the most widely used distance indicators,
type Ia supernovae (SNe Ia) show their importance in determining
cosmological parameters, which resulted in the discovery of the
accelerating expansion of the universe (\citealt{RIE98};
\citealt{PER99}). The result was exciting and suggested the
presence of dark energy. At present, SNe Ia are proposed to be
cosmological probes for testing the evolution of the dark energy
equation of state with time and testing the evolutionary history
of the universe (\citealt{RIESS07}; \citealt{KUZNETSOVA08};
\citealt{HOWEL09}). They were even chosen to check the consistency
of general relativity (\citealt{ZHAOGB10}). When SNe Ia are
applied as a distance indicator, the Phillips relation is adopted,
which is a linear relation between the absolute magnitude of SNe
Ia at maximum light and the magnitude drop of the B light curve
during the first 15 days following the maximum (\citealt{PHI93}).
This relation implies that the brightness of SNe Ia is mainly
determined by one parameter. It is generally agreed that the
amount of $^{\rm 56}$Ni formed during the supernova explosion
dominates the maximum luminosity of SNe Ia (\citealt{ARN82}), but
the origin of the different amount of $^{\rm 56}$Ni for different
SNe Ia is still unclear (\citealt{POD08}). Some numerical and
synthetical results showed that metallicity has an effect on the
final amount of $^{\rm 56}$Ni, and thus the maximum luminosity
(\citealt{TIM03}; \citealt{TRA05}; \citealt{POD06};
\citealt{BRAVO10}) and there do be some observational evidence of
the correlation between the properties of SNe Ia and metallicity
(\citealt{BB93}; \citealt{HAM96}; \citealt{WAN97};
\citealt{CAP97}; \citealt{SHA02}). However, the metallicity seems
not to have the ability to interpret the scatter of the maximum
luminosity of SNe Ia (\citealt{TIM03}; \citealt{GALLAGHER08};
\citealt{HOWEL09b}). \citet{NOM99, NOM03} suggested that the ratio
of carbon to oxygen (C/O) of a white dwarf at the moment of
explosion is the dominant parameter for the Phillips relation. The
higher the C/O, the larger the amount of nickel-56, and then the
higher the maximum luminosity of SNe Ia. The C/O ratio is a
function of the initial mass of the WD, which then is related to
the progenitor system of the SNe Ia. By comparing theory and
observations, the results of \citet{MENGXC09} and
\citet{MENGXC10a} upheld this suggestion. \citet{LESAFFRE06}
carried out a systematic study of the sensitivity of ignition
conditions for H-rich Chandra single degenerate exploders on
various properties of the progenitors, and suggested that the
central density of the WD at ignition may be the origin of the
Phillips relation (see also \citealt{POD08}). These authors
noticed that the more massive and/or the cooler the CO WD is when
accretion begins, the higher the central density is at ignition.
The central density is then also related to the progenitor system.
When one simulates the explosion of SNe Ia, the C/O and the
central density are always set to be free parameters
(\citealt{ROPKE06}). It is thus interesting to analyze whether
there is a correlation between the C/O and the central density. In
addition, analyzing how the initial mass of the CO WD and its
cooling time vary with the delay time is also an interesting task
(\citealt{GREGGIO10}). The purpose of this paper is to check these
interesting problems.

In section \ref{sect:2}, we describe our model. We show the
results in section \ref{sect:3} and give discussions and
conclusions in sections \ref{sect:4}.


\section{Model and Physics Inputs}\label{sect:2}
As suggested both \citet{NOM99, NOM03} and \citet{LESAFFRE06}, the
progenitor model is the so-called single degenerate model (SD),
i.e. the companion is probably a main sequence or a slightly
evolved star (WD+MS) or a red-giant (WD+RG) or a helium star (WD +
He star) (\citealt{WI73}; \citealt{NTY84}). The SD model is widely
accepted and studied by many authors (\citealt{YUN95};
\citealt{LI97}; \citealt{HAC99a,HAC99b}; \citealt{NOM99, NOM03};
\citealt{LAN00}; \citealt{HAN04, HAN06}; \citealt{CHENWC07,
CHENWC09}; \citealt{HAN08}; \citealt{LGL09};
\citealt{WANGB09a,WANGB09b}; \citealt{WANGB10};
\citealt{MENGXC10b,MENGXC10c}). In this paper, we also explore the
SD model. \citet{MENGXC10a} developed a comprehensive progenitor
model for SNe Ia. In the model, the mass-stripping effect by
optically thick wind (\citealt{HAC96}) and the effect of a
thermally unstable disk were included (\citealt{HKN08};
\citealt{XL09}). The prescription of \citet{HAC99a} for WDs
accreting hydrogen-rich material from their companions was applied
to calculate the WD mass growth. In \citet{MENGXC10a}, both the WD
+ MS and WD + RG scenarios are considered, i.e. Roche lobe
overflow (RLOF) begins at either the MS or the RG stage. After the
RLOF, the WD accretes hydrogen-rich material from its donor and
increases its mass smoothly. When the mass of the WD reaches 1.378
$M_{\odot}$ (\citealt{NTY84}), the WD is assumed to explode as an
SN Ia. They considered more than 1600 different WD close binary
evolution and obtained a parameter space for SNe Ia which is
summarized in an orbital period - secondary mass plane ($\log
P^{\rm i}, M_{\rm 2}^{\rm i}$). Our exploration is based on the
model of \citet{MENGXC10a}.

The C/O is a function of the initial WD mass of the progenitor
system, i.e. a higher WD mass leads to a lower C/O. We then use
the initial WD mass to represent the C/O. The central density is
mainly determined by the initial WD mass and its cooling time,
i.e. the more massive the WD and /or the longer the cooling time,
the higher the central density at ignition. The cooling time of
the CO WD is the time which elapses between its formation and the
start of the accretion phase. We thus use the initial WD mass and
its cooling time to represent the central density of a CO WD. The
following is how we obtain the initial WD mass and the cooling
time.

To obtain the distributions of the initial WD mass and its cooling
time, we carried out a series of Monte Carlo simulations via
Hurley's rapid binary evolution code (\citealt{HUR00, HUR02}). The
model grids obtained by \citet{MENGXC10a} were incorporated into
the code. We followed the evolution of $10^{\rm 7}$ binaries and a
circular orbit is assumed for all binaries. The metallicity is set
to solar metallicity, i.e. $Z=0.02$. The basic parameters for the
simulations are as follows: (1) a single star burst; (2) the
initial mass function (IMF) of \citet{MS79}; (3) the mass-ratio
distribution is constant; (4) the distribution of separations is
constant in $\log a$ for wide binaries, where $a$ is the orbital
separation; (5) the common envelope (CE) ejection efficiency
$\alpha_{\rm CE}$, which denotes the fraction of the released
orbital energy used to eject the CE, is set to be either 1.0 or
3.0 (see \citet{MENGXC10a} for details of the parameter input).

There are three channels that produce WD + MS systems and one
channel that produce a WD + RG system according to the situation
of the primary in a primordial system at the onset of the first
RLOF, i.e. the He star channel, the EAGB channel and the TPAGB
channel (see \citealt{MENGXC10a} for details about the channels).
The formed SD systems continue to evolve and the secondaries may
fill their Roche lobes at a later stage, leading to the start of
the RLOF. We assume that if the initial orbital period, $P_{\rm
orb}^{\rm i}$, and the initial secondary mass, $M_{\rm 2}^{\rm
i}$, of an SD system is located in the appropriate regions in the
($\log P^{\rm i}, M_{\rm 2}^{\rm i}$) plane for SNe Ia at the
onset of RLOF, an SN Ia is then produced. The cooling time is then
the difference between the formation time of an SD system and the
starting time of the RLOF.

\section{RESULTS}
\label{sect:3}

   \begin{figure}[h!!!]
   \centering
   \includegraphics[width=9.0cm, angle=270]{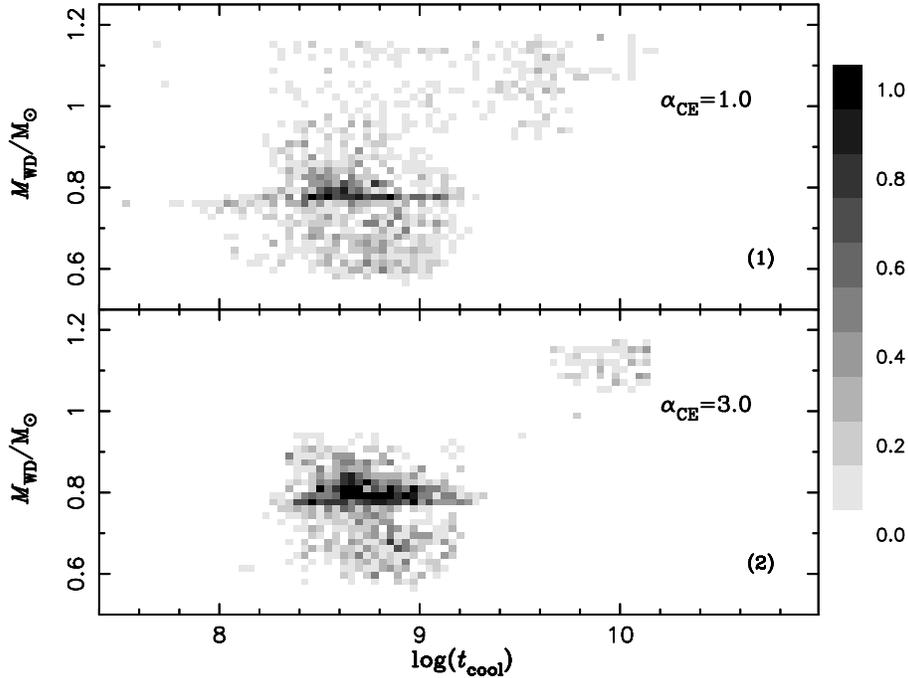}


   \caption{Distributions of initial WD masses and their cooling
   times
   for different $\alpha_{\rm CE}$.}
   \label{mwdcoolt}
   \end{figure}

\subsection{Distributions of the initial WD mass and their cooling time}\label{sect:3.1}

In Fig. \ref{mwdcoolt}, we show the distributions of the initial
WD mass and its cooling times for different $\alpha_{\rm CE}$. We
see from the figure that $\alpha_{\rm CE}$ does not significantly
affect the distributions. We do not find any correlation between
the initial WD mass and its cooling time, which means that taking
the C/O ratio and the central density of a CO WD as free
parameters is reasonable. However, the range of the WD mass
decreases and the mean value of the WD mass increases with the
cooling time, i.e. the range of the cooling time increases with
the WD mass (see Fig. 9 in \citealt{MENGXC10a}). In addition, it
is important to notice that there are some events with a very long
cooling time, i.e. several Gyrs, although the majority of CO WDs
have a cooling time shorter than 1 Gyr. The models in
\citet{LESAFFRE06} include cooling time ages of $\simeq1$ Gyr at
most. For a long cooling time, the CO WD may become more
degenerate before the start of the accretion phase, and some other
process like C and O separation or crystallization may occur, and
dominate the properties of the CO WD (\citealt{FONTAINE01}). The
simulation of the SNe Ia explosion under extremely degenerate
conditions should then be a more important problem than that
presented in literatures.

We also investigate the distributions of the initial orbital
period and the cooling time of a CO WD. Similar to the
distribution of the initial WD mass and its cooling time, no
correlation between the initial orbital period and the cooling
time of a CO WD was discovered.

   \begin{figure}
   \centering
   \includegraphics[width=90mm,angle=270.0]{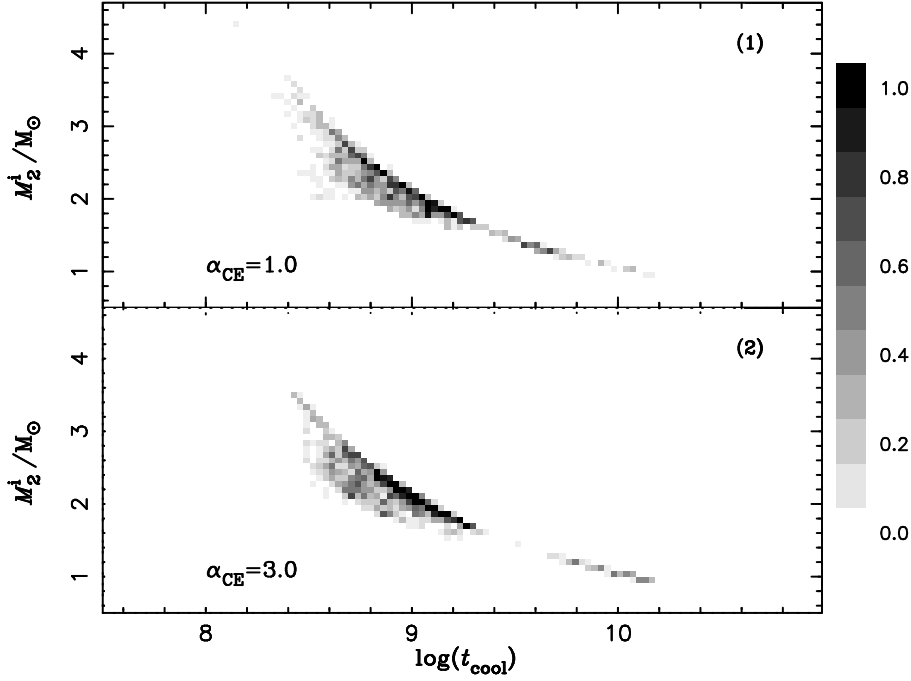}
   \caption{Distributions of the initial secondary mass and the
   cooling times of CO WDs for different $\alpha_{\rm CE}$.}
              \label{m2coolt}%
    \end{figure}
\subsection{Distributions of the initial secondary mass and the cooling time}\label{sect:3.2}
We present the distributions of the initial secondary mass and the
cooling time for different $\alpha_{\rm CE}$ values in Fig.
\ref{m2coolt}. Although $\alpha_{\rm CE}$ also does not
significantly affect the distributions, these distributions differ
remarkably from those of the initial WD mass and the cooling time.
The cooling time is highly relevant to the secondary mass. It is
clearly shown in Fig. \ref{m2coolt} that there is an upper limit
of the cooling time for a given secondary mass, and most of the
events have a cooling time equal to the upper limit. The upper
limit is derived from a constraint that the cooling time should be
smaller than the evolutionary age of the secondary. However, there
is still a scatter in the cooling time for a certain secondary
mass, which originated from the different initial orbital period
for an initial system with given WD and secondary masses. The
shorter the initial orbital period, the earlier the onset of the
RLOF between the WD and secondary, and then the shorter the
cooling time. In addition, the scatter decreases with the decrease
of secondary mass, and when the cooling time is larger than 2 Gyr,
the scatter almost disappears.

   \begin{figure}
   \centering
   \includegraphics[width=90mm,angle=270.0]{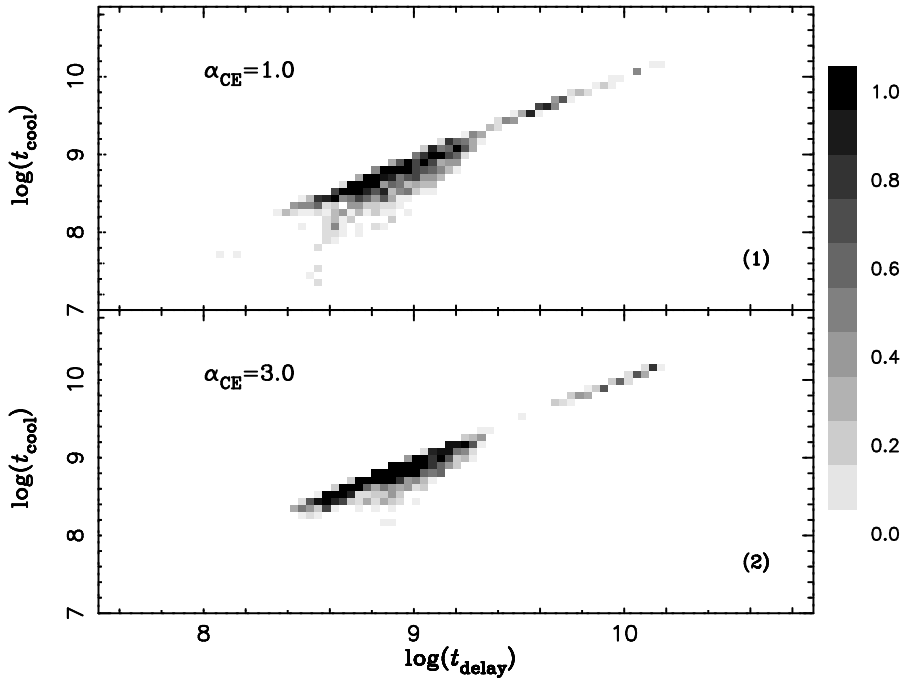}
   \caption{Relation between the
   cooling time of CO WDs and the delay time of SNe Ia for different $\alpha_{\rm CE}$.}
              \label{coolage}%
    \end{figure}
\subsection{Relation between the cooling time of CO WDs and the delay time of SNe Ia}\label{sect:3.3}
It is interesting to analyze how the WD cooling time varies with
the delay time of SNe Ia. In Fig. \ref{coolage}, the relation
between the cooling time and the delay time is presented. The
relation is very tight, and there is also an upper limit of the
cooling time for a given delay time. This result is directly from
the relation between the secondary mass and delay time. For most
of the progenitor systems with a certain secondary, the delay time
is the evolutionary lifetime of the secondary
(\citealt{MENGXC10a}; \citealt{GREGGIO10}). So, using the
secondary evolutionary time as the delay time of SNe Ia is an
excellent approximation when one applies an analytic method to
study the delay time of SNe Ia (\citealt{GREGGIO10}). There is
also a scatter for the relation between the cooling time and the
delay time, and the scatter decreases with the delay time. When
the delay time is larger than 2 Gyr, the cooling time is almost
equal to the delay time.

\section{DISCUSSION AND CONCLUSIONS}\label{sect:4}
In this paper, we do not find a correlation between the initial WD
mass and its cooling time. Since the C/O ratio is a function of
the initial WD mass and the central density for a WD with given
initial mass is mainly determined by the cooling time
(\citealt{NOM99, NOM03}; \citealt{LESAFFRE06}), our results may
imply that the C/O ratio and the central density at ignition are
free parameters when simulating an SNe Ia explosion. However, our
results still provide a constraint when simulating these types of
explosions, i.e. the cooling time of WDs with an initial mass less
than 1 $M_{\odot}$ is generally shorter than 1 Gyr, but it may be
as long as 15 Gyr for WDs with an initial mass larger than 1
$M_{\odot}$ (see Fig. \ref{mwdcoolt}). Because a high initial WD
mass means a lower C/O, and a massive WD and a long cooling time
leads to a high central density at ignition (\citealt{NOM99,
NOM03}; \citealt{LESAFFRE06}), the above result could imply that
WDs with a high C/O usually have a lower central density at
ignition, but those having the highest central density at ignition
generally have a lower C/O.

We also checked the effect of metallicity on the distributions of
the initial WD mass and the cooling time by $Z=0.001$
(\citealt{MENGXC10}) and no significant effect was found. Theory
and observations did confirm that the effect of metallicity cannot
explain the scatter in the maximum luminosity of SNe Ia
(\citealt{TIM03}; \citealt{GALLAGHER08}; \citealt{HOWEL09b}).
Metallicity should, at most, be the secondary parameter for the
Phillips relation, i.e. it is the origin for the scatter of the
Phillips relation (\citealt{POD06, POD08}). Furthermore, even
regarding the secondary parameter, there is not a consensus.
\citet{MAZZZALI06} suggested that a variation of the relative
content of ($^{\rm 54}$Fe+$^{\rm 58}$Ni) versus $^{\rm 56}$Ni may
be responsible for the observed scatter of the Phillips relation.
However, \citet{KASEN09} argued that the breaking of spherical
symmetry is a critical factor in determining both the Phillips
relation and the observed scatter around it. Then, the origin of
both the Phillips relation and the scatter of the relation still
should be investigated carefully.

We found that the range of the initial WD mass decreases and the
average WD mass increases with the cooling time, which is similar
to the relation between the initial WD mass and the delay time
found by \citet{MENGXC10a}. In addition, the scatter of the
cooling time also decreases with delay time. These results may
indicate that the difference among the WDs with s short cooling
time could be huge, but the properties of WDs with a long cooling
time might be more uniform (see from Fig. \ref{mwdcoolt} that all
the WDs with a cooling time longer than several Gyr have a mass
larger than 1.0 $M_{\odot}$). Since a long cooling time is
equivalent to a long delay time (see Fig. \ref{coolage}), our
results may imply that the properties of SNe Ia with long delay
times might be more uniform than those with short delay times. It
is widely known that there exists a scatter of the maximum
luminosity of SNe Ia, and the scatter is affected by its
environment. The most luminous SNe Ia always occur in spiral
galaxies, but both spiral and elliptical galaxies are hosts for
dimmer SNe Ia, which lead to a dimmer mean peak brightness in
elliptical than in spiral galaxies, i.e. the maximum luminosity of
SNe Ia hosted in elliptical galaxies is more uniform
(\citealt{HAM96}). These properties may be qualitatively
interpreted by the model developed by \citet{MENGXC10a}, at least
if the C/O ratio is the origin of the Phillips relation. However,
the quantitative study by \citet{ROPKE06} showed that the C/O has
only little impact on the amount of produced $^{\rm 56}$Ni. It
should be noticed that the results in \citet{ROPKE06} are model
dependent, and are much different from those of \citet{NOM99,
NOM03}. Considering that the central density is mainly determined
by the initial mass of the WD and its cooling time, the results in
Figs. \ref{mwdcoolt} and \ref{coolage}, i.e. the average values of
the WD mass and the cooling time increase with the delay time,
which means that the average of the central density increases with
the delay times of SNe Ia. A high central density at ignition
leads to a larger amount of the produced $^{\rm 56}$Ni
(\citealt{ROPKE06}). The effect of the central density on the
amount of $^{\rm 56}$Ni seems then to be opposite with
observations of \citet{HAM96}. Furthermore, the variation of the
amount of $^{\rm 56}$Ni derived from the central density only
amounts to about 7\%, which can not be used to interpret the
variation in the maximum luminosity of SNe Ia (\citealt{ROPKE06}).
Perhaps, the C/O ratio, the central density and the metallicity
all contribute to the variation of the maximum luminosity of SNe
Ia (\citealt{ROPKE06b}). Then, which is the dominant parameter for
the Phillips relation is still an open queation. In addition, as
the cooling/delay time increases, the CO WDs become more and more
degenerate, and even crystallization may occur
(\citealt{FONTAINE01}). What is the effect of the crystallization
on the amount of $^{\rm 56}$Ni should be an interesting problem.

 \normalem
\begin{acknowledgements}
This work was supported by Natural Science Foundation of China
under grant no. 10963001 and the Project of the Fundamental and
Frontier Research of Henan Province under grant no. 102300410223.
\end{acknowledgements}

\appendix                  

\section{This shows the use of appendix}

\label{lastpage}

\end{document}